\documentstyle [12pt,a4p,epsfig,amsmath,multicol]{article}
\begin{document}
\vspace*{0.6cm}

\begin{center} 
{\normalsize\bf A comment on the paper `Coherence in Neutrino Oscillations'
  by C.Giunti. Either lepton flavour eigenstates or neutrino oscillations
  do not exist}
\end{center}
\vspace*{0.6cm}
\centerline{\footnotesize J.H.Field}
\baselineskip=13pt
\centerline{\footnotesize\it D\'{e}partement de Physique Nucl\'{e}aire et 
 Corpusculaire, Universit\'{e} de Gen\`{e}ve}
\baselineskip=12pt
\centerline{\footnotesize\it 24, quai Ernest-Ansermet CH-1211Gen\`{e}ve 4. }
\centerline{\footnotesize E-mail: john.field@cern.ch}
\baselineskip=13pt
 
\vspace*{0.9cm}
\abstract{ The definition of `lepton flavour eigenstates' introduced in a
 recent paper hep-ph/0302045 is in disagreement with the Standard Model
 prediction for the structure of the leptonic charged current and implies
 the absence of neutrino oscillations following pion decay, in contradiction
 to experimental data}
\vspace*{0.9cm}
\normalsize\baselineskip=15pt
\setcounter{footnote}{0}
\renewcommand{\thefootnote}{\alph{footnote}}
\newline
PACS 03.65.Bz, 14.60.Pq, 14.60.Lm, 13.20.Cz 
\newline
{\it Keywords ;} Quantum Mechanics,
Neutrino Oscillations.
\newline

\vspace*{0.4cm}

  
  In a recent short note~\cite{Giunti}, Giunti has claimed to refute the claim
  of Reference~\cite{JHF1} that experimental measurements of the decay width ratio
  in pion decay:
 \[R_{e/\mu} = \Gamma(\pi^- \rightarrow e^- \overline{\nu})/
 \Gamma(\pi^- \rightarrow \mu^- \overline{\nu})\] and of the elements of the 
  MNS~\cite{MNS} lepton flavour/mass mixing matrix exclude the possiblity,
  for massive neutrinos, 
  of the production of a coherent `lepton flavour eigenstates' $\overline{\nu}_e$,
  $\overline{\nu}_{\mu}$ in pion decay. The discussion of Reference~\cite{JHF1}
   is first briefly recalled, before considering Giunti's counter argument.
   \par In the Standard Electroweak Model the invariant amplitude for the pion 
    decay process: $\pi^- \rightarrow \ell^- \overline{\nu}_i$ where $\ell =e,\mu$
   and $\overline{\nu}_i$ denotes a mass eigenstate, is:
 \begin{equation}
  {\cal M}_{\ell i} = \frac{G}{\sqrt{2}}f_{\pi}m_{\pi}V_{ud}
     \overline{\psi}_{\ell}(1-\gamma_5)U_{\ell i}\psi_{\overline{\nu}_i}
   \equiv {\cal M}_{\ell i}^D U_{\ell i}   
  \end{equation}
  where $G$ is the Fermi constant, $f_{\pi}$ and $m_{\pi}$ are the 
  pion decay constant and mass respectively,  and
  $V_{ud}$, $U_{\ell i}$ are, respectively, elements of the CKM~\cite{CKM}
   quark flavour/mass mixing
  matrix and the MNS lepton flavour/mass mixing matrix.
  For simplicity, only two flavour mixing is considered so that $i =1,2$ and
   $U_{\ell i}$ can be taken to be real.
   \par The standard discussion of neutrino oscillations following pion
   decay starts with the assumption that the state produced is the coherent
  superposition, $\psi_{\overline{\nu}_{\ell}}$, of mass eigenstates 
   defined by:
  \begin{equation}
   \psi_{\overline{\nu}_{\ell}}  \equiv  U_{\ell 1}\psi_{\overline{\nu}_1}+
 U_{\ell 2}\psi_{\overline{\nu}_2}
 \end{equation}
  Denoting the coherent state in Eqn(2) as
  $\psi_{\overline{\nu}_{\ell}}(\tau = 0)$, where $\tau$ is the anti-neutrino proper
 time, the standard neutrino oscillation phase is derived by introducing time
 evolution of the mass eigenstates in Eqn(2), according to the Schr\"{o}dinger 
 equation in the anti-neutrino rest frames, as, for example,
  in Reference~\cite{KaysPDG}:
  \begin{equation}
   \psi_{\overline{\nu}_{\ell}}(\tau) = U_{\ell 1}\psi_{\overline{\nu}_1}(0)e^{-im_1 \tau}
 +U_{\ell 2}\psi_{\overline{\nu}_2}(0)e^{-im_2 \tau}
 \end{equation}
 As is well known, the assumption that the anti-neutrinos have equal momentum, $p$, but 
  different energies, and are ultra-rlativisitic, then leads to the  standard
  result for the oscillation phase: $\phi = (m_1^2-m_2^2)L/2p$ where $L$ is the 
   source-detector distance. It is clear, then, that the assumption that the
   coherent state $ \psi_{\overline{\nu}_{\ell}}$ of Eqn(2) is produced in pion
  decay is crucial for the derivation of the standard neutrino oscillation
  phase. Combining now Eqns(1) and (2) allows the invariant amplitude
  ${\cal M}_{\ell \ell}$ for the production of the state $\psi_{\overline{\nu}_{\ell}}$
  to be introduced:
 \begin{equation}
  {\cal M}_{\ell \ell} \equiv {\cal M}_{\ell 1}+  {\cal M}_{\ell 2}
   = \frac{G}{\sqrt{2}}f_{\pi}m_{\pi}V_{ud}
     \overline{\psi}_{\ell}(1-\gamma_5) \psi_{\overline{\nu}_{\ell}}   
  \end{equation}
  This amplitude {\it does produce} the state $\psi_{\overline{\nu}_{\ell}}$
  in pion decay, and, as shown above, subsequent observation of the anti-neutrinos
  via charged current interactions will give rise to neutrino oscillations 
  with the standard phase. Another immediate consequence of Eqn(4) is the 
  prediction:
  \begin{equation}
  R_{e/\mu} \equiv \frac{\Gamma(\pi^- \rightarrow e^- \overline{\nu}_e)}
 {\Gamma(\pi^- \rightarrow \mu^- \overline{\nu}_{\mu})} =
  \left(\frac{m_e}{m_{\mu}}\right)^2 \left [\frac{m_{\pi}^2-m_e^2}
   {m_{\pi}^2-m_{\mu}^2}\right]^2
  \left(\frac{U_{e 1}+U_{e 2}}
   {U_{\mu 1}+U_{\mu 2}}\right)^2
  \end{equation}
  where it is assumed that purely kinematical effects of non-vanishing 
  anti-neutrino masses may be neglected in both the invariant amplitudes and the phase 
  space factors. As shown in Reference~\cite{JHF1} the prediction, Eqn(5),
  is excluded with a huge statistical significance by the experimental
   measurements of $R_{e/\mu}$~\cite{PDG} and the MNS matrix elements
   ~\cite{GGN}. It must then be concluded that the initial hypothesis
    underlying Eqn(5), that the coherent state in Eqn(2) is produced in pion
    decay, is false.
    \par In Reference~\cite{Giunti} it is proposed to use, instead of Eqn(2),
    a different definition of the `lepton flavour eigenstate',
    $\psi_{\overline{\nu}_{\ell}}^G$, according to the definitions:
  \begin{equation}
  {\cal M}_{\ell \ell}^G \equiv {\cal M}_{\ell 1} U_{1 \ell}+
    {\cal M}_{\ell 2} U_{2 \ell}
 = \frac{G}{\sqrt{2}}f_{\pi}m_{\pi}V_{ud}
     \overline{\psi}_{\ell}(1-\gamma_5) \psi_{\overline{\nu}_{\ell}}^G   
  \end{equation}     
  and 
  \begin{eqnarray}
   \psi_{\overline{\nu}_{\ell}}^G &\equiv&  U_{\ell 1} U_{1 \ell}\psi_{\overline{\nu}_1}+
 U_{\ell 2} U_{2 \ell} \psi_{\overline{\nu}_2} \nonumber \\
   & = & |U_{\ell 1}|^2 \psi_{\overline{\nu}_1}+|U_{\ell 2}|^2\psi_{\overline{\nu}_2}
 \nonumber \\
    & \simeq & (|U_{\ell 1}|^2+|U_{\ell 2}|^2)\psi_{\overline{\nu}_0} = 
   \psi_{\overline{\nu}_0}
 \end{eqnarray}
 where $\psi_{\overline{\nu}_0}$ denotes the wavefunction of a massless anti-neutrino,
  the unitarity of the MNS matrix: $ |U_{\ell 1}|^2+|U_{\ell 2}|^2 = 1$, has been
 used, and, as in Eqn(5), the purely kinematical effects of non-vanishing anti-neutrino
  masses are neglected. Since the MNS elements do not appear in Eqn(6),
  the prediction given by this equation for $R_{e/\mu}$ is the same as the text-book
  massless anti-neutrino result, which is in excellent agreement with experiment
  and provides no information on the values of the MNS elements. However, since
  the amplitude (6) evidently does not produce the coherent state
   $\psi_{\overline{\nu}_\ell}$ of Eqn(2), it predicts the absence of 
   neutrino oscillations following pion decay. This seems now to be experimentally
  excluded by the observation of such oscillations in both atmospheric
  neutrinos~\cite{KajTot} and the K2K~\cite{K2K} experiment.

   \par To summarise: the analysis of Reference~\cite{JHF1} shows that the production
   of `lepton flavour eigenstates' as in Eqn(2) in the final state of pion decay is 
   experimentally excluded, but that neutrino oscillations are possible, albeit with
  a phase that is often quite different from the standard one. The analysis of Reference
  ~\cite{Giunti} avoids the constraint of Eqn(5), but predicts the absence of 
   of neutrino oscillations following pion decay, in contradiction to the results
   of both the  Superkamiokande and K2K experiments.

    \par Actually, the {\it anstatz} of Eqn(6), which appears to have been 
   constructed  precisely to avoid the constraint imposed by Eqn(5), is without
   any physical justification. As first pointed out by Shrock~\cite{Shrock1}
   and as recently recalled by Pallin and Snellman~\cite{PS}, the correct Standard
   Model structure of the leptonic charged current in pion
   decay is just that given by Eqns(1) and (2)
  above, not by Eqns(6) and (7).
    The `lepton flavour eigenstate'
    of Eqn(2), though useful as a formal device to write the Lagrangian or Hamiltonian
    of the charged weak current in a compact form (as, for example, in Eqn(4) of
     Reference~\cite{Giunti}
    or Eqn(2) of Reference~\cite{PS}) does not occur in the amplitude of any physical
    process. The final and intial states in the amplitudes of all physical processes
    in the Standard Model
    are mass eigenstates. This is true of all fermions, whether quarks or leptons.
    A close analogy, in the quark sector, to pion decay is provided by 
    the decay of W bosons into quarks~\cite{JHF2}:
     \begin{equation} 
  \rm{W}^- \rightarrow  \overline{\rm{c}} \rm{d}~~~~~~~~~\rm{W}^- \rightarrow
   \overline{\rm{c}} \rm{s},~~~~~~~
     \rm{W}^- \rightarrow  \overline{\rm{c}} \rm{b}
     \end{equation}
      The analogue of the neutrino `lepton flavour eigenstate' of Eqn(2) would be 
      the `charm flavour eigenstate of d-type quarks':
  \begin{equation}
   \psi_c  \equiv  V_{c d}\psi_d +
 V_{c s}\psi_s + V_{c b}\psi_b 
 \end{equation}
 Of course, such a `coherent' state has no relevance to W decays. The decays into the
   different quark flavours shown above occur incoherently, that is in different
   physical processes.
  In the Standard Model the different neutrino
  flavours (mass eigenstates) are, just like the different quark flavours 
   just discussed, also produced
  {\it incoherently}~\cite{Shrock1} in different physical processes. Because of this
   incoherent
   behaviour the ratio $R_{e/\mu}$ is indeed quite insensitive to the values of
   the MNS matrix elements. This is because:
  \begin{eqnarray}
   \Gamma(\pi^+ \rightarrow \ell^+\overline{\nu}) & = & 
    \Gamma(\pi^+ \rightarrow \ell^+\overline{\nu}_1)+ 
     \Gamma(\pi^+ \rightarrow \ell^+\overline{\nu}_2) \nonumber \\
   & \propto & |{\cal M}_{\ell 1}^D|^2 |U_{\ell 1}|^2+
      |{\cal M}_{\ell 2}^D|^2 |U_{\ell 2}|^2  \nonumber \\
     & \simeq &  |{\cal M}_{\ell 0}^D|^2(|U_{\ell 1}|^2+|U_{\ell 2}|^2) =
      |{\cal M}_{\ell 0}^D|^2      \ 
 \end{eqnarray}
  where, unlike in Eqn(7), the unitarity of the MNS matrix has been used in 
  a physically correct manner. In fact, Eqn(10) leads to the same prediction
   for $ R_{e/\mu}$ as Eqns(6) and (7): there is no sensitivity to the values of the MNS
   matrix elements. This is, however, a simple consequence of the incoherent nature of
   the production processes as exemplified in the first line of Eqn(8), not of the
   unphysical {\it anstatz} of Eqns(6) and (7). 
   \par Because the different neutrino mass eigenstates are produced incoherently,
   in different physical processes, they are not necessarily produced at the
   same time (as they must be if they are produced in the coherent state of Eqn(2))
   so that the propagator of the source particle gives an important contribution to the
   oscillation phase, which is then found to be process dependent and, in many cases,
   markedly different from the standard one~\cite{JHF1,JHF2}. 
 
\pagebreak

\end{document}